\documentclass[twocolumn,showpacs,preprintnumbers,amsmath,amssymb,APSl,prd,nofootinbib,superscriptaddress]{revtex4-1}

\usepackage{dcolumn}
\usepackage{bm}
\usepackage{ifpdf}
\usepackage{hyperref}
\usepackage{dcolumn}
\usepackage{bm}
\usepackage[spanish,english]{babel}
\usepackage{amsfonts}
\usepackage{amssymb}
\usepackage{graphicx}
\usepackage[latin1]{inputenc}

\newcommand{\LL}{\mathcal{L}}

\newcommand{\be}{\begin{equation}}
\newcommand{\en}{\end{equation}}
\newcommand{\bea}{\begin{eqnarray}}
\newcommand{\ena}{\end{eqnarray}}

\begin{document}

\title{Microscopic wormholes and the geometry of entanglement}

\author{Francisco S. N.~Lobo}\email{flobo@cii.fc.ul.pt}
\affiliation{Centro de Astronomia e Astrof\'{\i}sica da
Universidade de Lisboa, Campo Grande, Ed. C8 1749-016 Lisboa,
Portugal}
\author{Gonzalo J. Olmo} \email{gonzalo.olmo@csic.es}
\affiliation{Departamento de F\'{i}sica Te\'{o}rica and IFIC, Centro Mixto Universidad de
Valencia - CSIC. Universidad de Valencia, Burjassot-46100, Valencia, Spain}
\author{D. Rubiera-Garcia} \email{drubiera@fisica.ufpb.br}
\affiliation{Departamento de F\'isica, Universidade Federal da
Para\'\i ba, 58051-900 Jo\~ao Pessoa, Para\'\i ba, Brazil}


\date{\today}

\begin{abstract}

It has recently been suggested that Einstein-Rosen (ER) bridges can be interpreted as maximally entangled states of two black holes that form a complex Einstein-Podolsky-Rosen (EPR) pair. This relationship has been dubbed as the $ER = EPR$ correlation. In this work, we consider the latter conjecture in the context of quadratic Palatini theory. 
An important result, which stems from the underlying assumptions about the geometry on which the theory is constructed, is the fact that all the charged solutions of the quadratic Palatini theory possess a wormhole structure.
Our results show that spacetime may have a foam-like microstructure with wormholes generated by fluctuations of the quantum vacuum. This involves the spontaneous creation/annihilation of entangled
particle-antiparticle pairs, existing in a maximally entangled state connected by a non-traversable wormhole. Since the particles are produced from the vacuum and therefore exist in a singlet state, they are necessarily entangled with one another. This gives further support to the $ER=EPR$ claim.

\end{abstract}

\maketitle

{\it Introduction.} In the context of the firewall debate \cite{Almheiri:2012rt, Almheiri:2013hfa}  (see \cite{BPZ} for earlier work), it has recently been argued that Einstein-Podolsky-Rosen (EPR) correlations \cite{EPR} and Einstein-Rosen (ER) bridges \cite{ER} are actually related \cite{Maldacena:2013xja}. For instance, the ER bridge between two black holes is created by EPR-like correlations between the microstates of the two black holes \cite{Maldacena:2013xja}. This conjecture has been schematically dubbed the $ER = EPR$ correlation. More specifically, the ER bridge is a special kind of EPR correlation, where the entanglement has a geometric manifestation. Despite the fact that the two black holes exist in separate and non-interacting spacetimes, their geometry is connected by an ER bridge, and the entanglement is represented by identifying the bifurcate horizons \cite{Maldacena:2013xja}. Note that the ER bridge construction is due to the fact that specific coordinate systems naturally cover the two asymptotically flat regions of maximally extended spacetimes \cite{ER}, and the key ingredient of the bridge construction is the existence of an event horizon. Thus, although the geometry is connected through the bridge/tunnel, the two exterior geometries are not in causal contact and information cannot be transmitted across the bridge \cite{fullwheel}. This is an essential point to be consistent with the fact that entanglement does not imply non-local signal propagation in order to conserve causality.

Recently, the $ER=EPR$ claim received further support through specific models. Indeed, it was shown that the holographic dual of two colored quasiparticles can be constructed in maximally supersymmetric Yang-Mills theory entangled in a color singlet EPR pair \cite{Jensen:2013ora, Chernicoff:2013iga}. In the holographic dual the entanglement is encoded in a geometry of a non-traversable wormhole on the worldsheet of the flux tube connecting the pair. In this context, it was also pointed out that the proposed bulk dual of an entangled quark--anti-quark pair described above \cite{Jensen:2013ora} corresponds to the Lorentzian continuation of the tunneling instanton describing a Schwinger pair creation in the dual field theory \cite{Sonner:2013mba}. This observation supports and further explains the claim in \cite{Jensen:2013ora} that the bulk dual of an EPR pair is a string with a wormhole on its world sheet. It was also suggested that this constitutes an AdS/CFT realization of the creation of a Wheeler wormhole \cite{Wheeler}.

Wheeler used the source-free Maxwell equations, coupled to Einstein gravity, with the seasoning of nontrivial topology, to build models for classical electrical charges and all other particle-like entities in classical physics \cite{Wheeler}. This analysis culminated in the ``geon'' concept, coined by Wheeler to denote a ``gravitational-electromagnetic entity''. Building on this pioneering work, Misner and Wheeler, in 1957, presented a tour de force wherein Riemannian geometry of manifolds of nontrivial topology was investigated with an ambitious view to explaining all of physics \cite{Misner:1957mt}. Indeed, it has also been argued that if topology change is allowed in quantum gravity, it is possible to create a Wheeler wormhole \cite{GarStrom}. In fact, Maldacena and Susskind take the radical position in that the $ER=EPR$ correlations are inseparably linked in a theory of quantum gravity, even for systems of entangled particles. The intimate relationship between spacetime geometry and the underlying degrees of freedom of entanglement has also been explored in the literature \cite{lit_entangle}.

In this work, we consider a concrete $ER=EPR$ context, namely, in the geon solutions arising in quadratic Palatini gravity. In fact, the underlying assumptions about the geometry on which the theory is constructed implies that all the charged solutions of the quadratic Palatini theory  possess a wormhole structure.
Our results support the view that spacetime could have a foam-like microstructure with wormholes generated by fluctuations of the quantum vacuum involving the spontaneous creation/annihilation of entangled particle-antiparticle pairs. This gives further support to the $ER=EPR$ conjecture.

{\it Palatini Ricci-squared theory.} Consider a gravity theory coupled to matter with the action
\begin{equation}\label{eq:action}
S[g,\Gamma,\psi_m]=\int d^4x \sqrt{-g} \left[ \LL_G +\LL_m(g,\psi_m)  \right]  \ ,
\end{equation}
where $\LL_G=f(R,Q)/(2\kappa^2)$ represents the gravitational Lagrangian, $\kappa^2$ being a constant with suitable dimensions [in General Relativity (GR), $\kappa^2 \equiv 8\pi G$], and with the definitions $R=g^{\mu\nu}R_{\mu\nu}$, $Q=g^{\mu\alpha}g^{\nu\beta}R_{\mu\nu}R_{\alpha\beta}$, $R_{\mu\nu}={R^\rho}_{\mu\rho\nu}$, where the Riemann tensor is constructed with the connection $\Gamma \equiv \Gamma^{\lambda}_{\mu\nu}$, i.e.,
\begin{equation}\label{eq:Riemann}
{R^\alpha}_{\beta\mu\nu}=\partial_{\mu}
\Gamma^{\alpha}_{\nu\beta}-\partial_{\nu}
\Gamma^{\alpha}_{\mu\beta}+\Gamma^{\alpha}_{\mu\lambda}\Gamma^{\lambda}_{\nu\beta}-\Gamma^{\alpha}_{\nu\lambda}\Gamma^{\lambda}_{\mu\beta} \,.
\end{equation}
$\LL_m(g,\psi_m)$ represents the matter Lagrangian density, which is minimally coupled to the spacetime metric $g_{\mu\nu}$ and $\psi_m$ collectively denotes the matter fields, to be specified below.

Now, the Palatini approach assumes that the connection $\Gamma_{\mu\nu}^{\lambda}$, which defines the affine structure, is {\it a priori} independent of the metric, so that the two sets of field equations are obtained from the variation of the action (\ref{eq:action}) with respect to the metric and the connection as
\begin{eqnarray}
f_R R_{\mu\nu}-\frac{f}{2}g_{\mu\nu}+2f_QR_{\mu\alpha}{R^\alpha}_\nu &=& \kappa^2 T_{\mu\nu}\label{eq:met-varX}\\
\nabla_{\beta}^\Gamma\left[\sqrt{-g}\left(f_R g^{\mu\nu}+2f_Q R^{\mu\nu}\right)\right]&=&0  \ ,
 \label{eq:con-varX}
\end{eqnarray}
respectively (vanishing torsion and $R_{[\mu\nu]}=0$ have been assumed, see \cite{or13b}), where $T_{\mu\nu}$ is the energy-momentum tensor of the matter, and the notation $f_X \equiv df/dX$ has been used. By means of algebraic manipulations \cite{Barragan2010, or12a}, Eq.(\ref{eq:con-varX}) can be written as $\nabla_{\beta}^\Gamma[\sqrt{-h} h^{\mu\nu}]=0 $, which implies that $\Gamma^\alpha_{\beta\gamma}$ is the Levi-Civita connection of an auxiliary metric $h_{\mu\nu}$ defined as
\begin{equation} \label{eq:h-g}
h^{\mu\nu}=\frac{g^{\mu\alpha}{\Sigma_{\alpha}}^\nu}{\sqrt{\det \hat{\Sigma}}} \ , \quad
h_{\mu\nu}=\left(\sqrt{\det \hat{\Sigma}}\right){{{\Sigma}^{-1}}_{\mu}}^{\alpha}g_{\alpha\nu} \,.
\end{equation}
The matrix $\hat\Sigma$ is defined by ${\Sigma_\alpha}^{\nu}\equiv \left(f_R \delta_{\alpha}^{\nu} +2f_Q {P_\alpha}^{\nu}\right)$, where ${P_\mu}^\nu\equiv R_{\mu\alpha}g^{\alpha\nu}$. The field equations imply that ${P_\mu}^\nu$, $R$, $Q$, and $\hat\Sigma$ are algebraic functions of the matter fields. Using Eqs. (\ref{eq:h-g}), we can write Eq. (\ref{eq:met-varX}) as \cite{Barragan2010, or12a, lor13}
\begin{equation} \label{eq:fieldequations}
{R_{\mu}}^{\nu}(h)=\frac{\kappa^2}{\sqrt{\det \hat{\Sigma}}}\left(\LL_G\delta_{\mu}^{\nu}+  {T_\mu}^{\nu} \right) \ .
\end{equation}
Given that $\LL_G$ and $\hat\Sigma$ are functions of ${T_\mu}^{\nu}$, we see that $h_{\mu\nu}$ satisfies a set of GR-like second-order field equations and, since $h_{\mu\nu}$ and $g_{\mu\nu}$ are algebraically related, it follows that $g_{\mu\nu}$ also verifies second-order equations. From Eqs. (\ref{eq:h-g}) and (\ref{eq:fieldequations}) it is easily seen that the vacuum field equations (${T_\mu}^{\nu}=0$) of the Palatini theory (\ref{eq:action}) recover the vacuum general relativistic equations \cite{lor13}, with possibly a cosmological constant (depending on the specific form of the gravitational Lagrangian $\LL_G$), which is a general property of Palatini gravities \cite{Ferraris,or13b}.  Thus, no new propagating degrees of freedom are present, and these theories are free of the ghost-like instabilities \cite{Deser,Banados,Olmo:2013gqa} which appear in the metric formulation of theories with non-linear powers of the Ricci tensor.

In the following, we consider the specific case of a dynamical spherically symmetric spacetime perturbed by an ingoing null flux of energy and electric charge. The pressureless flux of ingoing charged matter has a stress-energy tensor given by $T_{\mu\nu}^{\rm flux}= \rho_{\rm in} k_\mu k_\nu$, where $\rho_{\rm in}$ is the energy density of the ingoing flux, and $k_{\mu}$ is a null vector $k_{\mu}k^{\mu}=0$.
Consider now a line element of the form
\begin{equation}\label{eq:ds2g}
ds^2=-A(x,v) e^{2\psi(x,v)}dv^2+ 2e^{\psi(x,v)}dv dx+r^2(v,x)d\Omega^2 \,,
\end{equation}
in which the integration of the Maxwell equations, $\nabla_\mu F^{\mu\nu}=4\pi J^\nu$, where $J^\nu\equiv \Omega(v) k^\nu$ is the current of the ingoing flux, lead to $r^2 e^{\psi(x,v)}F^{xv}=q(v)$, where $q(v)$ is an integration function and $\Omega(v)\equiv q_v/4\pi r^2$.

As a working hypothesis, we choose a specific quadratic extension of GR, given by
\begin{equation}\label{eq:quadratic}
f(R,Q)=R+l_P^2(a R^2+Q) \ ,
\end{equation}
where $l_P^2 \equiv \hbar G/c^3$ is Planck's length squared and $a$ a free parameter. The motivation of this model stems from the fact that quadratic curvature corrections arise in the quantization of fields in curved spacetime \cite{PT}, in approaches to quantum gravity based both in string theory \cite{strings} and loop quantum gravity \cite{o08, Bertolami:2013uwl}, and when GR is regarded as an effective theory of quantum gravity \cite{Cembranos}.
 
Taking into account the matter sources and the gravity Lagrangian, one finds $R=0$ and $Q=\kappa^2 q^4/4\pi r^8$. After some lengthy, but straightforward calculations, the field equations (\ref{eq:fieldequations}) are finally written as
\begin{equation}\label{eq:Rmn}
{R_{\mu}}^{\nu}(h)=\left(
\begin{array}{cccc}
-\frac{{\kappa}^2q^2(v)}{8\pi r^4\sigma_+} & \frac{e^{-\psi}\kappa^2\rho_{in}}{\sigma_+\sigma_-}  & 0 & 0  \\
0 & -\frac{{\kappa}^2q^2(v)}{8\pi r^4\sigma_+} & 0 & 0  \\
0& 0& \frac{{\kappa}^2q^2(v)}{8\pi r^4\sigma_-} & 0 \\
0& 0& 0 & \frac{{\kappa}^2q^2(v)}{8\pi r^4\sigma_-}
\end{array}
\right)  \ ,
\end{equation}
where  $\sigma_{\pm}\equiv 1\pm {\kappa}^2 l_P^2 q^2(v)/(4\pi r^4)$.

Now, the solution to the field equations (\ref{eq:Rmn}), in terms of the physical metric $g_{\mu\nu}$, is given by the following line element (for more details of these calculations, we refer the reader to \cite{Lobo:2013vga})
\begin{eqnarray}
ds^2&=&-\left[\frac{1}{\sigma_+}\left(1-\frac{1+\delta_1 (v) G(z)}{\delta_2(v) z \sigma_{-}^{1/2}}\right)- \frac{2l_P^2 \kappa^2 \rho_{\rm in}}{\sigma_{-}(1-\frac{2r_c^4}{r^4})}\right]dv^2
    \nonumber\\
 &&+ \frac{2}{\sigma_+}dvdx+r^2(x,v) d\Omega^2 \,, \label{eq:ds2final}
\end{eqnarray}
where
\begin{equation}
r^2(x,v)=\frac{1}{2}\left [x^2+\sqrt{x^4+4 r_c^4(v)} \right]\,, \label{eq:rtx0}
\end{equation}
with $r_c(v)\equiv\sqrt{r_q(v)l_P}$ and $r_q^2(v)\equiv \kappa^2 q^2(v)/4\pi$.
The factors $\delta_1(v)=\frac{1}{2r_S(v)} \sqrt{r_q^3(v)/l_P}$ and $\delta_2(v)\equiv r_c(v)/r_S(v)$ have been defined for simplicity, with $z(x,v)\equiv r(x,v)/r_c(v)$, $r_S(v) \equiv 2M(v)$ (where $M(v)$ is a mass function), the function $G(z)$ satisfies $G_z=(z^4+1)/[z^4 \sqrt{z^4-1}]$, and we have used the relation $dr/dx=\sigma_{-}^{1/2}/\sigma_{+}$ (at constant $v$), which is deduced from Eqs. (\ref{eq:h-g}).

Consider now a compact charged perturbation propagating within the interval $[v_i,v_f]$ in an initial flat Minkowski space. From Eq. (\ref{eq:rtx0}), in the interval $v<v_i$ we have $r^2(x,v)=x^2$, which extends from zero to infinity. Entering the $v\ge v_i$ region, this radial function, which measures the area of the
$2$-spheres of constant $x$ and $v$, never becomes smaller than $r_c^2 (v)$, with the minimum located at $x=0$. If we now consider the region $v>v_f$, in which $\rho_{\rm in}$ is again zero, the result is a static geometry.
One can verify \cite{or12a} that in this static geometry curvature scalars generically diverge at $x=0$ except if the charge-to-mass ratio $\delta_1^f \equiv \delta_1(v\ge v_f)$ takes the value  $\delta_1^f=\delta_1^* \simeq 0.572$. This $\delta_1^*$  is a constant that appears in the series expansion of $G(z)=-1/\delta_1^*+2\sqrt{z-1}+\ldots$ as $z\to 1$. The smoothness of the geometry when $\delta_1^f=\delta_1^*$, together with the fact that $r(x)$ reaches a minimum at $x=0$, naturally justifies the extension of the domain of $x$ to the negative real axis, thus showing that the area function $r^2(x)$ bounces off to infinity as $x\to -\infty$. This puts forward the existence of a wormhole structure supported by the electric field with its throat located at $x=0$. The minimum of $r^2(x,v)$ together with the fact that the electric flux per surface unit at $x=0$, given by $\Phi/4\pi r_c^2(v)=q(v)/r_c^2(v)=\sqrt{c^7/2\hbar G^2}$, is a constant independent of $q(v)$ and $M(v)$ confirms that the wormhole (topological) structure exists even when $\delta_1^f\neq\delta_1^*$, i.e., we always have two sides ($x\in ]-\infty,+\infty[$) regardless of the possible existence of (local) curvature divergences at $x=0$.

For $|x|\gg r_c(v)$, the line element (\ref{eq:ds2final}) quickly recovers the behavior found in GR in both the dynamic and the static case. In fact, it is a simple matter to show that as $l_P \rightarrow 0$ in Eqs. (\ref{eq:quadratic}) and (\ref{eq:Rmn}) we recover the GR limit, and the theory yields the well known Bonnor-Vaidya solution of GR \cite{BV}. Event horizons are thus expected in general, and their location is almost coincident with the GR prediction (within $O\left(r_q^2 l_P^2/r^4\right)$ corrections) for not too small black holes.

To better understand the geometry around the wormhole in the final static configurations, consider an expansion of the metric component $g_{vv}$ in (\ref{eq:ds2final})   around $r/r_c\equiv z\approx 1$ \cite{lor13}
\begin{equation}
g_{vv}= \frac{\left(1-\delta _1/\delta_1^*\right)}{4\delta _2 \sqrt{z-1}}-\frac{1}{2}\left(1-\frac{\delta _1}{ \delta _2}\right)+O(\sqrt{z-1}) \label{eq:gtt_series} \ .
\end{equation}
Confronting this expansion with a numerical analysis shows that  when $\delta_1=\delta_1^*$ the sign of the term $\left(1-{\delta_1^*}/{\delta _2 }\right)$ in (\ref{eq:gtt_series}) determines whether an event horizon exists or not \cite{or12a}. Since  $\delta_1^*/\delta_2=r_q/2l_P$, it follows that the event horizon is absent if $r_q<2l_P$. This inequality can be written as a constraint on the charge of the system. In fact, expressing the charge as $q=N_q e$, where $e$ is the electron charge and $N_q$ the number of charges,  one finds that $r_q=2l_P N_q /N_q^c$, where $N_q^c\equiv \sqrt{2/\alpha_{\rm em}} \approx 16.55$, which leads to $\delta_1^*/\delta_2\equiv N_q/N_q^c$. Therefore, when $\delta_1=\delta_1^*$, an event horizon exists if  $N_q>N_q^c$ (and its location agrees well with GR for $N_q \gtrsim 2N_q^c$). It can also be shown \cite{or12a} that for $\delta_1 < \delta_1^*$ there is always an event horizon (Schwarzschild-like case), while for $\delta_1 > \delta_1^*$ we may have one, two, or no horizons (Reissner-Nordstrom-like case).

For static charged configurations, therefore, non-traversable wormhole solutions exist. These solutions consist of an electric flux going through one of the sides of the wormhole  and coming out through the other side, thus generating a spherically symmetric electric field \cite{lor13} associated to a negative charge on one side and a positive charge on the other. We emphasize that an electric field of this kind does not require the existence of sources for its generation, as first shown by Wheeler and Misner \cite{Misner:1957mt}. For all practical purposes, there is no difference between this kind of charge, arising from a pure electric field trapped in the nontrivial topology (going through a wormhole), and a standard point-like charge. The mass of these solutions is also naturally associated with the energy stored in the electric field, which is regularized to a finite value due to the bound $r\ge r_c$.

The results presented above support the view that spacetime could have a foam-like microstructure with wormholes generated by fluctuations of the quantum vacuum involving the spontaneous creation/annihilation of entangled particle-pairs (the antiparticle-particle pair in a maximally entangled singlet state is connected by a wormhole). This gives further support to the $ER=EPR$ conjecture.

{\it Geometry of entanglement.} Maldacena and Susskind \cite{Maldacena:2013xja} argue that the geometry of the ER bridge is the geometric manifestation of the entanglement between the two black holes. They further consider several hypothetical scenarios in that an entangled black hole pair may arise through EPR pair productions. A more subtle scenario involves the formation of one-sided black hole by gravitational collapse.  Assume that an outside observer, over a long period of time, captures the Hawking radiation emitted by the black hole, and then collapses the respective radiation into their own black hole. This involves the entanglement between the original Hawking radiation and the one-sided black hole. However, it is argued that, in general, entanglement should be associated with wormhole formation, in that Hawking radiation should be connected to the black hole interior through micro-wormholes, which encode the entanglement. In fact, the collapse of the Hawking radiation into a second black hole entails that the micro-wormholes combine to form a macroscopic ER bridge between the two black holes \cite{Maldacena:2013xja}.

In this work, we have provided an explicit realization of this microscopic idea but with the geon solutions arising in the quadratic Palatini theory outlined above. The solutions found arise in charged pairs of $+N_q$ and $-N_q$, and are space-like separated when the wormhole lies behind an event horizon, which are the non-traversable Wheeler wormholes that form the geometric entanglement between the two black holes. These can be created through charged perturbations of ingoing fluxes of energy, described above, or by intense magnetic fields \cite{GarStrom, lor14}. However, the important point to bear in mind, is that they are explicitly connected by a non-traversable Planckian wormhole of area $A_{WH}=4\pi r_c^2=(8\pi l_P^2)N_q/N_q^c$.

Consider a one-sided black hole, formed by gravitational collapse and a set of entangled pairs $\{+N_{q_i},-N_{q_i}\}$. Assume now that the black hole captures all the positive charges. These charges add up to form an object of charge $Q_{\rm tot}=\Sigma_i N_{q_i}$, which contains a wormhole of area $A_{\rm tot} \propto N_{Q_{\rm tot}}$. This larger wormhole is connected to the $\{-N_{q_i}\}$ partners through their respective microscopic tunnels. Note that the area of each of these microscopic tunnels is $A_i\propto N_{q_i}$ and that the total sum of those areas coincides with the area of the larger tunnel $A_{\rm tot}$. This property is nontrivial and is a result of the linear growth of the area of the wormhole with the charge. As a result of this process, a number of bridges exist between the black hole and the external partners of the entangled pairs. This means that the black hole is now entangled with external particles but is not maximally entangled. The latter maximal entanglement only occurs between the two members of the entangled pairs.

Relative to the issue of entanglement transfer, consider now two pairs of entangled charged particles, $\{N_{q_1},-N_{q_1}\}$ and $\{N_{q_2},-N_{q_2}\}$, where each pair is connected by a microscopic wormhole. There is no {\it a priori} entanglement between the pairs $1$ and $2$. Assume now that $N_{q,1}$ and $-N_{q,2}$ are absorbed by an isolated (one-sided) black hole and a stationary state is achieved. In this new state, the black hole has an internal wormhole of area $A\propto |N_{q_1}-N_{q_2}|$ and, therefore, is entangled with the external partners $-N_{q_1}$ and $+N_{q_2}$. Note that in the process of formation of the new state, the electric flux of $N_{q_1}$ entered through $-N_{q_2}$ and reached $N_{q_2}$, which at the same time reaches $-N_{q_1}$ to close the circuit. This means that the capture of $N_{q_1}$ and $-N_{q_2}$ has transferred a certain amount of entanglement between the external members. The black hole is thus entangled with $-N_{q_1}$ and $+N_{q_2}$ and those elements between them. This process of entanglement transfer is naturally visualized in terms of propagation of flow through microscopic wormholes. Note that by combining different charges we can transfer and/or destroy entanglement by local operations but cannot create it.

{\it Discussion.} The fact that all the charged solutions of the quadratic Palatini theory  possess a wormhole structure is an important result which stems from the underlying assumptions about the geometry on which the theory is constructed.
Our approach allows independent variations of the metric and affine degrees of freedom and, as a result, the field equations determine the form of the connection, i.e., we do not set it {\it a priori} to be the Levi-Civita connection of the metric. Allowing the connection to account for the non-metric properties of the geometry, we find that the theory replaces the usual point-like curvature singularities found in GR by wormholes, which can be formed dynamically. If cubic or quartic curvature corrections are considered in the action, like in the Born-Infeld gravity theory proposed by Deser and Gibbons \cite{Deser,Banados}, these wormhole solutions persist unaltered if one follows a Palatini approach \cite{Olmo:2013gqa}, which gives robustness to our results.
These nontrivial topological structures support the view that spacetime could have an underlying foam-like microstructure with wormholes generated by fluctuations of the quantum vacuum. The latter involves the spontaneous creation/annihilation of particle-antiparticle pairs, which exist in a maximally entangled state connected by a non-traversable wormhole. We have found that by capturing elements of an entangled pair, black holes can become entangled with external systems and transfer and/or destroy entanglement. This entanglement is directly supported by a microscopic wormhole, as hypothesized in \cite{Maldacena:2013xja}, which gives further support to the $ER=EPR$ claim.

{\it Acknowledgments.}
FSNL acknowledges financial support of the Funda\c{c}\~{a}o para a Ci\^{e}ncia e Tecnologia through an Investigador FCT Research contract, with reference IF/00859/2012, funded by FCT/MCTES (Portugal), and grants CERN/FP/123615/2011 and CERN/FP/123618/2011. GJO is supported by the Spanish grant FIS2011-29813-C02-02, the Consolider Program CPANPHY-
1205388, the JAE-doc program of the Spanish Research Council (CSIC), and the i-LINK0780 grant of CSIC. DRG is supported by CNPq (Brazilian agency) through project No. 561069/2010-7. This work has also been supported by CNPq through project No. 301137/2014-5. The authors are indebted to A. Plastino for useful discussions.


\begin{thebibliography}{99}

\bibitem{Almheiri:2012rt}
  A.~Almheiri, D.~Marolf, J.~Polchinski, and J.~Sully,
  JHEP {\bf 1302}, 062 (2013).

\bibitem{Almheiri:2013hfa}
  A.~Almheiri, D.~Marolf, J.~Polchinski, D.~Stanford, and J.~Sully,
  JHEP {\bf 1309}, 018 (2013).

\bibitem{BPZ} 
S. L. Braunstein, S. Pirandola and K. Zyczkowski, 
Phys. Rev. Lett. 110, 101301 (2013).

\bibitem{EPR}
A. Einstein, B. Podolsky, and N. Rosen,
Phys.\ Rev.\ {\bf 47}, 777 (1935).


\bibitem{ER}
  A.~Einstein and N.~Rosen,
  Phys.\ Rev.\  {\bf 48}, 73 (1935).

\bibitem{Maldacena:2013xja}
  J.~Maldacena and L.~Susskind,
  arXiv:1306.0533 [hep-th].


\bibitem{fullwheel}
  R.~W.~Fuller and J.~A.~Wheeler,
  Phys.\ Rev.\  {\bf 128}, 919 (1962).

\bibitem{Jensen:2013ora}
  K.~Jensen and A.~Karch,
  Phys.\ Rev.\ Lett.\  {\bf 111}, 211602 (2013).

\bibitem{Chernicoff:2013iga} 
 M.~Chernicoff, A.~Güijosa and J.~F.~Pedraza, 
 JHEP {\bf 1310}, 211 (2013). 

\bibitem{Sonner:2013mba}
  J.~Sonner,
  Phys.\ Rev.\ Lett.\  {\bf 111}, 211603 (2013).

\bibitem{Wheeler}
J. A. Wheeler, Phys. Rev. \textbf{97}, 511 (1955).

\bibitem{Misner:1957mt}
  C.~W.~Misner and J.~A.~Wheeler,
  Annals Phys.\  {\bf 2}, 525 (1957).

\bibitem{GarStrom}
  D.~Garfinkle and A.~Strominger,
  Phys.\ Lett.\ B {\bf 256}, 146 (1991).

\bibitem{lit_entangle}
  S.~Ryu and T.~Takayanagi,
  Phys.\ Rev.\ Lett.\  {\bf 96}, 181602 (2006);
%
  M.~Van Raamsdonk,
  Gen.\ Rel.\ Grav.\  {\bf 42}, 2323 (2010)
  [Int.\ J.\ Mod.\ Phys.\ D {\bf 19}, 2429 (2010)].

\bibitem{or13b}	G. J. Olmo and D. Rubiera-Garcia, Phys. Rev. D \textbf{88} 084030 (2013).

\bibitem{Barragan2010} C. Barragan and G. J. Olmo, Phys. Rev. D \textbf{82}, 084015 (2010);
C. Barragan, G. J. Olmo, and H. Sanchis-Alepuz, Phys. Rev. D \textbf{80}, 024016 (2009).

\bibitem{or12a}
G. J. Olmo and D. Rubiera-Garcia, Phys. Rev. D \textbf{86}, 044014 (2012);
Int. J. Mod. Phys. D \textbf{21}, 1250067 (2012);
Eur. Phys. J. C \textbf{72}, 2098 (2012).

\bibitem{lor13}
F.~S.~N.~Lobo, G.~J.~Olmo, and D.~Rubiera-Garcia,
  JCAP {\bf 07}, 011 (2013).

\bibitem{Ferraris} M. Ferraris, M. Francaviglia, and I. Volovich,  Class. Quant. Grav. \textbf{11}, 1505 (1994); A. Borowiec, M. Ferraris, M. Francaviglia, and I. Volovich, Class. Quant. Grav. \textbf{15}, 43 (1998).

\bibitem{Olmo:2013gqa}
G.~J.~Olmo, D.~Rubiera-Garcia and H.~Sanchis-Alepuz,
  Eur.\ Phys.\ J.\ C {\bf 74}, 2804 (2014).

\bibitem{Deser} S. Deser and G. W. Gibbons, Class. Quant. Grav. {\bf 15}, L35 (1998).

\bibitem{Banados} M. Ba\~nados and P. G. Ferreira, Phys. Rev. Lett. {\bf 105}, 011101 (2010).

\bibitem{PT} L. Parker and D. J. Toms, \emph{Quantum field theory in curved space-time: quantized fields and gravity} (Cambridge University Press U. K., 2009); N. D. Birrel and P. C. W. Davies, \emph{Quantum fields in curved space} (Cambridge University Press, U. K., 1982).

\bibitem{strings} M. Green, J. Schwarz, and E. Witten, \emph{Superstring theory} (Cambridge University Press, U. K., 1987); T. Ortin, \emph{Gravity and strings} (Cambridge University Press, U. K. 2004).

\bibitem{o08}
G.~J.~Olmo and P.~Singh,
  JCAP \textbf{01}, 030 (2009).

\bibitem{Bertolami:2013uwl} 
O.~Bertolami and J.~Páramos,  Phys.\ Rev.\ D {\bf 89} 044012 (2014).

\bibitem{Cembranos} J. A. R. Cembranos, Phys. Rev. Lett. \textbf{102}, 141301 (2009).

\bibitem{Lobo:2013vga}
F.~S.~N.~Lobo, J.~Martinez-Asencio, G.~J.~Olmo and D.~Rubiera-Garcia, Phys.\ Lett.\ B {\bf 731} 
16 (2014). 

\bibitem{BV} W. B. Bonnor and P. C. Vaidya, Gen. Rel. Grav. \textbf{1}, 127 (1970).


\bibitem{lor14} F. S. N. Lobo, G. J. Olmo, and D. Rubiera-Garcia, in preparation.


\end{thebibliography}
\end{document}